\documentclass[proceedings]{JHEP}
\usepackage{amsfonts}

\catcode`\@=11 
\def\numberbysection{\@addtoreset{equation}{section} 
        \def\theequation{\thesection.\arabic{equation}}} 
 
\numberbysection 
 
\newcommand{\beq}{\begin{equation}} 
\newcommand{\eeq}{\end{equation}} 
\newcommand{\bea}{\begin{eqnarray}} 
\newcommand{\eea}{\end{eqnarray}} 
 
\def\s{\sigma} 
\def\a{\alpha} 
\def\b{\beta} 
\def\d{\delta}
 
\def\eps{\epsilon}

\def\l{\lambda}

\def\r{\rho}

\def\Im{\hbox{\rm Im\ }} 
 
\def\nl{\nonumber\\} 
\def\bra{\langle} 
\def\ket{\rangle} 
\def\TTT{\bra TTT\ket}

\conference{Non-perturbative Quantum Effects 2000}
 
\title{On the c-theorem in more than two dimensions
\thanks{Preprint SPhT t00/131; proceedings of the TMR Conference
``Non-perturbative Quantum Effects 2000'', Paris, 7-13 Sept. 2000,
D. Bernard and B. Julia Eds., to be published in the J. of High-Energy
Physics.}
}

\author{
A. Cappelli${}^a$, G. D'Appollonio${}^a$, R. Guida${}^b$ and 
N. Magnoli${}^c$
\\
${}^a$ I.N.F.N. and Dipartimento di Fisica, 
Largo E. Fermi 2, I-50125 Firenze, Italy,
\\ 
${}^b$ CEA-Saclay, Service de Physique Th\'eorique,
F-91191 Gif-sur-Yvette, France,
\\
${}^c$ I.N.F.N. and Dipartimento di Fisica, Via Dodecaneso 33, 
I-16146 Genova, Italy.
}

\abstract{ 
Several pieces of evidence have been recently brought up 
in favour of the $c$-theorem in four and higher dimensions,
but a solid proof is still lacking.
We present two basic results which could be useful for this search: 
{\it i)} the values of the putative $c$-number for free field theories
in any even dimension, which illustrate some properties of this number;
{\it ii)} the general form of three-point function of the stress tensor
in four dimensions, which shows some physical consequences of
the $c$-number and of the other trace-anomaly numbers.
}

\begin{document}  
 
\section{Introduction}

\subsection{The $c$-theorem in two dimensions}

The renormalization-group (RG) flow is defined as 
the one-parameter motion in the space of (renormalized) coupling constants
$\{g^i,\ i=1,2,\dots\}$,
\beq
\frac{\rm d}{{\rm d}t} \equiv 
-\ \b^i(g)\ \frac{\partial}{\partial g^i}\ ,
\label{rg-flow}\eeq
with ``velocities'' given by the beta-functions;
the flow corresponds to a change of scale in the field theory
which grows towards the infrared.

The Zamolodchikov $c$-theorem\cite{cth} holds for
unitary, renormalizable quantum field theories in two dimensions;
it says that there exists a positive-definite real function 
of the coupling constants $c(g)$ such that: 

\noindent {\it i)}
it is monotonically decreasing along the flow,
\beq
\frac{\rm d}{{\rm d}t}\ c \le \ 0 \ ;
\label{mono}\eeq

\noindent {\it ii)}
it is stationary at the fixed points $g^i=(g^*)^i$,
\beq
\b^i(g^*)\ = \ 0 \ \ \leftrightarrow\ \ 
\frac{\partial}{\partial g^i}\ c(g)\Bigg\vert_{g^*} =\ 0\ ;
\label{stat}\eeq

\noindent {\it iii)}
at the fixed points, it equals the Virasoro central charge $c$ of the
corresponding conformal field theory,
\beq
c\left( g^*\right) \ = \ c\ .
\label{cstar}\eeq

This theorem implies some fundamental properties of the RG flow:

\noindent {\it i)}
The flow necessarily ends into fixed points (or fixed surfaces); 
there cannot exist
limit cycles or strange attractors, which are other possible 
asymptotic behaviours for the solutions of
non-linear differential equations (\ref{rg-flow}).

\noindent {\it ii)}
The fixed points are classified according to the value
of their central charge; we can think the space of theories\footnote{
Namely, the space of coupling constants.}
as a mountain landscape, with
the fixed points located at the tips, the saddles and the 
 valleys bottoms.

The central charge is a measure of the ``number of degrees
of freedom'' and its decreasing along the RG flow can be
viewed as the consequence of ``coarse graining'', the
integration of high-energy degrees of freedom
in the Wilsonian approach to the renormalization group
\cite{kada}. 
Note that a theory with an asymptotic limit cycle would
have a never-ending infrared flow, with degrees
of freedom periodically dying out and coming back;
it would be very difficult to make sense of this RG behaviour
in a unitary field theory.
In conclusion, the $c$-theorem confirms our intuitive understanding 
of the RG flow.

Let us remark some aspects of the Zamolodchikov proof that will
be useful for the forecoming discussion:

\noindent {\it i)}
The inputs of the proof are just ``kinematics'',
i.e. general properties of Poincare invariance,
unitarity and renormalizability -- there are no hypotheses on the
dynamics of the theory.

\noindent {\it ii)}
The function $c(g)$ is finite once the coupling
constants are renormalized; moreover, its critical
value is uniquely defined, since the trace anomaly is
both finite and universal (as any other anomaly).
It follows that $c(g)$ is defined globally 
(i.e. non-perturbatively) on the whole space of theories.


\subsection{The $c$-theorem in higher dimension: motivations
and overview of the results}

The consequences of the $c$-theorem on the RG flow
are so general that it is natural to expect its
extension to higher-dimensional field theories.
However, more than ten years have passed since the
first attempts to a generalization \cite{cardy}\cite{jo}\cite{cfl}.
First of all, a straightforward extension of the 
Zamolodchikov argument is not possible \cite{cfl}. 
Secondly, in odd dimension $d=3,5,\cdots$, the
$c(g)$-function lacks the natural global definition given by 
the trace anomaly, because the latter is equal to zero
(actually, it is very easy to construct functions
which are monotonically decreasing along the flow
but are discontinuous at fixed points).

It seems that the extension of the $c$-theorem
to higher dimension requires a new ingredient, 
possibly involving the field-theory dynamics.
In this respect, we believe that the eventual proof 
of the theorem could teach us new properties of 
field theories and of their interaction with gravity
(through their stress tensor).
Therefore, the interest of the $c$-theorem extends
beyond the proof of mandatory properties of the RG flow.

Several works have recently discussed
the $c$-theorem in four dimensions,
by providing new arguments for the proof \cite{fl}\cite{ath}
\cite{c-a} and by analysing examples of RG flows for 
the trace anomaly coefficients $a$, $c$ and $a'$ \cite{afgj}\cite{fgpw}. 
These are defined by the following expression\footnote{
The coefficient $\l=- 1/(2880\cdot 4\pi^2)$ is included
to normalize the values of $a$ and $c$ to one for
the free conformal-invariant scalar field.
} \cite{duff}:
\beq
\bra T_\mu^\mu \ket \ =\ \l \left(
a\ E \ - \ 3c \ W \ +\ a' \ D^2 {\cal R} \right)\ ,
\label{trace}\eeq
where $\int\sqrt{g}\ E = \chi$ is the Euler characteristics,
$W$ is the square of the Weyl tensor and
${\cal R}$ is the curvature scalar.

\medskip
\noindent {\bf $a$-theorem:} \\
$a_{UV} > a_{IR}\ $ has been exactly proven \cite{afgj}
for the non-trivial RG flows among $N=1$ 
supersymmetric gauge theories found by Seiberg, most notably those
in the ``conformal window'' \cite{c-berg}. 

\medskip
\noindent{\bf $c$-theorem:} \\
$c_{UV} > c_{IR}\ $ cannot be true in general, because 
counterexamples are known \cite{cfl}\cite{afgj}; however, it holds for
the field theories which have a gravity dual theory
according to the AdS/CFT correspondence \cite{hs}\cite{fgpw},
such as the $N=4$ supersymmetric gauge theories.
In these theories, the ratio $c/ a$ is an overall 
fixed constant, thus these results also support
the $a$-theorem. 
The AdS/CFT correspondence has provided a lot of evidence 
for the irreversibility of the RG flow and a proof
of the theorem has been found in this context \cite{fgpw}
(see also Ref.\cite{c-a}).

\medskip
\noindent{\bf $a'$-theorem:} \\
the decreasing of $a'$ along the flow can be easily
proven, but this does not imply the theorem,
because the function $a'(g)$ cannot be globally defined 
in the space of theories \cite{clv}.
Actually, $a'$ is not well defined at fixed points, 
because it corresponds to a scheme-dependent term in the trace
anomaly (\ref{trace}): $D^2 {\cal R}$ is the Weyl 
variation of the local term $\int\sqrt{g}{\cal R}^2$
in the effective action \cite{pbb}.
This problem can be cured by assuming a proportionality
between $a'$ and $a$, as proposed in Ref.\cite{ath};
however, this amounts to a 
strong dynamical hypothesis on the effective action.

In conclusion, the most promising formulation of the
theorem in four dimensions involves the coefficient $a$ of the 
Euler term in the trace anomaly 
(as first suggested in Ref. \cite{cardy}).


\section{The anomaly $a$ as a measure of degrees of freedom}

In this Section, we suppose as a working hypothesis
that the $a$-theorem is true in any even dimension
$d \ge 4$ (the Euler term is always present in the trace
anomaly \cite{ds}); we want to understand
how the $a$-number is actually measuring the
number of degrees of freedom in field theory.
To this extent, it is interesting to compute the
value of $a$ in several free theories and study its
dependence on spin and dimension \cite{ad}.

We consider the free conformal invariant theories of the scalar
$(S)$, Dirac fermion $(F)$ and antisymmetric tensor
$(AT)$ fields\footnote{This is the $p$-form field,
with $p=(d-2)/2$ for conformal invariance at the 
classical level \cite{ad}.} and compute their trace
anomalies on the $d$-dimensional sphere $S^d$. 
We use the well-known zeta-function regularization of the 
Euclidean partition function, given by the determinant of 
the Laplacian $\Delta$ (the Hodge-de Rahm operator \cite{ad}) 
acting on the respective fields.
Under a scale transformation of the metric,
$g_{\mu\nu} \to \exp(2\a)\  g_{\mu\nu}$,
the variation of the partition function is:
\beq
\frac{\rm d}{{\rm d}\a} \log Z\left[ S^d \right] =
\zeta_{\Delta}\left(s=0\right) 
= 2\int_{S^d}\ \bra T_\mu^\mu \ket\ ,
\label{inta}\eeq
where
\beq
\zeta_{\Delta}\left(s\right) = \sum_n\ \frac{1}{\l^s_n}\ ,
\label{z-fun}\eeq
is the zeta function associated to the Laplacian, whose
eigenvalues are denoted by $\l_n$. 

\TABLE[t]{
\begin{tabular}{|c||c|c|c|c|c|c|c|c|}
\hline 
$d  $ & $4$  &$ 6 $  &$ 8 $  & $10 $  &$ 12 $  & $14$  &    &$ 2k$   \\ 
\hline\hline
$a(S)$ & $1$  &$ 1$  & $  1$  &  $ 1 $  &  $ 1$  &  $ 1$  &   &      \\ 
\hline
$a(F)$ & $11$ &$ \frac{191}{5}$ &$\frac{2497}{23}$  &$\frac{73985}{263} $  
 &$\frac{92427157}{133787}$   & $\frac{257184319}{157009} $ &$ \cdots$ & \\ 
\hline
$a(AT)$ &$62$ &$\frac{3978}{5}$ &$\frac{161020}{23}$ &$\frac{13396610}{263} $  
&$\frac{44166621324}{133787} $ &$\frac{310708060404}{157009}$ &$ \cdots$ & \\ 
\hline\hline
$r(S) $ &$ 1$ & $1$ & $ 1$  & $  1 $  &  $ 1$   & $  1 $ &  &$ 1$      \\ 
\hline
$r(F)$ &$ 2.75$ & $4.77$ &$ 6.79$ &  $ 8.79$& $ 10.79$& $12.80$&$\cdots$ &
$ \simeq 2k$\ \ \ \\ 
\hline
$r(AT$) &$ 31 $ &$ 132.6$ &$ 350.0 $&$ 727.7 $&$ 1310.$ &$2142.$ &
$\cdots$ &$ \simeq (2k)^3$ \\ 
\hline
\end{tabular}
\caption{Values of  $a$ and of the weight per field 
component $r$ in various even dimensions $d$, with asymptotic
behaviours for $d\to\infty$.}
\label{atab}
}

The trace anomaly in any even dimension contains
the Euler term we are interested in, plus a number
of terms which are Weyl-covariant polynomials of the
Weyl tensor and its derivatives \cite{ds}.
These additional terms vanish on the geometry of
the sphere, which is related to Euclidean space
by a Weyl transformation; therefore, the trace
anomaly (\ref{inta}) is completely given by
the Euler term.

Thus, we can write the following equation for
the trace anomaly on any conformally-flat space ${\cal M}$:
\beq
\frac{\rm d}{{\rm d}\a} \log Z\left[ {\cal M} \right] =
2\l\ a\ \chi\left({\cal M} \right) = 
\frac{\zeta_{\Delta}(0)}{2} \ \chi\left({\cal M} \right)\ ,
\label{a-val}\eeq
where $\chi$ is the Euler characteristic
in $d$ dimensions ($\chi\left(S^d\right)=2$),
$\l$ is the normalization constant for $a$ and 
$\zeta_\Delta(0)$ is computed on $S^d\ $ \cite{ad}.

Equation (\ref{a-val}) determines the values of $a$
once the normalization $\l$ is chosen.
We first consider the 
normalization $\l=1$ (call $a=\widehat{a}$ in this case);
this is a rather natural choice, because
 $\widehat{a}$ becomes the proportionality constant
between two universal pure numbers which are 
$d$ and scale independent: a topological number on the r.h.s. of (\ref{a-val})
and the regularized number of modes of the Laplacian on the l.h.s.
(which can also be thought of as the number of ``effective zero modes''
\cite{ad}).

The anomaly number $\widehat{a}(\s)$ 
divided by the number of field components $ n(\s)$,
 $\s=S,F,AT$, is found to decrease with the dimension 
and to vanish in the limit $d=\infty$ \cite{ad}:
\beq
\frac{\widehat{a}(\s)}{n(\s)}\ \to \ 0\ ,
\quad {\rm for} \ d\to\infty\ ,
\label{a-lim}\eeq
with
\bea
&& n(S)=1\ , \qquad n(F)=2^{d/2}\ , \nl
&& n(AT) = {(d-2)!\over \left[\left(\frac{d}{2}-1 \right)!\right]^2}\ .
\label{n-val}\eea
The behaviour of $\widehat{a}$ is consistent with the known fact
that these free theories 
become semiclassical in the limit of large dimensionality\footnote{
This can be seen by putting the theories on a space-time lattice.}:
the anomaly is a quantum effect and should go
to zero (once properly normalized).

We now discuss the use of $a$ as a measure
of degrees of freedom in the spirit of the
$c$-theorem; we should use another normalization $\l=\l(d)$
in (\ref{a-val}), such that the
scalar field is counted the same value in any dimension, say:
\beq
a(B)\equiv 1\ , \qquad {\rm any}\ d\ .
\label{a-def}\eeq
(This determines the value of $\l=\l(4)$ in (\ref{trace})).
The values of $a$ in this normalization are reported in Table \ref{atab}
together with the ratios per field component, 
\beq
r(\s)\equiv \frac{a(\s)}{n(\s)} \ .
\label{r-def}\eeq

We find that the ratios do not approach $\ 1\ $ for
large $d$, but actually grow like $O(d)$ and $O(d^3)$
for the fermion and antisymmetric tensor fields, respectively.
The measure of degrees of freedom 
given by $a$ is very different from the classical value
$n(\s)$, even when the theories become semiclassical; 
 the higher-spin fields are weighted much more
than the lower-spin ones, as is already apparent in $d=4$.
This is the main result of the work \cite{ad}.
The same qualitative enhancement is found \cite{ad} for the other coefficient
$c$ in the trace anomaly (\ref{trace}) and 
for the gravitational chiral anomaly \cite{lagw}.

This result for the $a$-counting is rather counter-intuitive but does not
directly imply an obstruction for
the $a$-theorem in higher dimensions: it
does not lead to contradictions in the RG flows checked so far.
It is a peculiar behaviour that one should 
keep in mind for further investigations of the $a$-theorem.


\section{The three stress-tensor correlator in four dimensions.}

The previous discussions have shown the important role
of the trace anomaly in the various attempts to
extend the $c$-theorem above two dimensions.
In this respect, 
a better understanding of the physical consequences of the trace
anomaly is very useful.
Since the $a$ and $c$ coefficients of the $d=4$ trace anomaly
(\ref{trace}) are scheme-independent quantities,
it is possible to relate them to finite, 
scheme-inde\-pen\-dent amplitudes of the stress-tensor correlators
and thus to physical quantities in flat space \cite{kaguma}.

\subsection{Two-dimensional preliminaries}

Let us first review the relation of the trace anomaly to two-dimensional
correlators and its key role for the dispersive proof of the $c$-theorem
\cite{cfl}.
In two dimensions, the two-point correlator of the stress tensor
can be written in momentum space as follows: 
\bea
\!\! &&\bra T_{\mu\nu}(p)\ T_{\r\s}(-p) \ket \nl
\!\! && = \frac{\pi}{3}\ A(p^2)
\left(p_\mu p_\nu - \d_{\mu\nu} \right)
\left(p_\r p_\s - \d_{\r\s} \right) ,
\label{2t2}\eea
where the form of the tensor structure is required by conservation,
i.e. by Diffeomorphism invariance.
The dimensional analysis shows that the scalar
amplitude $A(p^2)$ has dimension $(-2)$ and therefore
is finite in perturbation theory and scheme independent.
At fixed points, it becomes:
\beq
A(p^2) \ =\ \frac{c}{p^2} \ , \qquad\qquad {\rm fixed\ points}.
\label{2d-c}
\eeq
This Equation gives the desired relation of 
the anomaly coefficient $c$ with the scheme-independent
correlator, which plays an important role in the
conformal field theory \cite{id}
($\bra T(z)T(0) \ket=c/2z^4$ in coordinate space).

Off criticality, the amplitude satisfies the dispersion relation:
\bea
A(p^2)&=&\int ds\ \frac{\r(s)}{s+p^2}\ ,\nl
\r(s)&=& \frac{1}{\pi}\ \Im A\left( p^2 = -s \right)\ .
\label{disp}\eea
In this Equation, $\r(s) d s$ is a positive-definite
dimensionless spectral measure \cite{cfl}, whose critical limit is:
\beq
\r(s) \to\ c\ \d (s)\ ,\qquad\qquad   {\rm fixed\ points}.
\label{rho-c}\eeq
Using this measure, one can obtain another proof
of the Zamolodchikov theorem, as follows \cite{cfl}:
off-criticality, the measure contains a delta term plus a
smooth positive function peaked at $s=m^2$, where $m$
is the typical mass scale of the theory:
\beq 
\r (s)= c_0 \d (s) + \r_{\rm smooth}\left(s/m^2 \right)\ .
\label{rho-off}\eeq
In the infrared limit $m\to\infty$, the peak 
will move to infinity and the smooth function will go to
zero in a weak sense (i.e. as a distribution);
therefore, the coefficient of the remaining delta-function
is identified with the central charge of the infrared theory:
$\ c_0=c_{IR}$. 
On the other hand, in the ultraviolet limit $m\to\ 0$,
the smooth function should go to a delta function which
adds up to the first term of (\ref{rho-off}), such that the total integral
gives the central charge of the ultra-violet theory: 
$\int_0^\infty \r(s)ds = c_{UV}$.

These properties of the spectral measure imply the
following sum rule, which is an equivalent form of the $c$-theorem:
\bea
c_{UV}-c_{IR} &=& \int_\eps^\infty\ ds\ \r(s) \nl
&=& \frac{3}{4\pi}\int_{|x|>\eps} d^2x\ x^2
\bra T_\mu^\mu(x) T_\nu^\nu (0) \ket \nl
&>& \ 0\ .
\label{s-rule}\eea
In this Equation, we also wrote the
expression of the sum rule in coordinate space \cite{sumrule}.

The previous analysis can be extended to four dimensions 
\cite{cfl}\cite{kaguma} starting from:
\bea
&&\bra T_{\mu\nu}(p)\ T_{\r\s}(-p) \ket \nl
&&= A_0(p^2) \ {\cal P}^{(0)}_{\mu\nu,\r\s}
+ A_2(p^2) \ {\cal P}^{(2)}_{\mu\nu,\r\s}\ ,
\label{2t4}\eea
where ${\cal P}^{(0)}$ is the polynome in (\ref{2t2})
and ${\cal P}^{(2)}$ is an analogous expression 
projecting on spin-two intermediate states.
The two amplitudes $( A_0 ,\ A_2)$ now have  
zero dimension, thus they are superficially
divergent and scheme dependent; their critical limits are:
\bea
A_0(p^2) &\ \to\ &\l \ a'\ ,\qquad\qquad {\rm fixed\ points},\nl
A_2(p^2) &\ \to\ &\l \frac{c}{4}\ \log 
\left( \frac{p^2}{\mu^2} \right) \ ,
\label{a-crit}\eea
where $a'$ is the coefficient of the scheme-dependent
term in the trace anomaly (\ref{trace}) and
$\mu$ is the renormalization scale.
These expressions explicitly show the scheme dependences:
$A_i(p^2)\to$ $ A_i(p^2) + {\rm const.}$; it follows
that the two-point function, although positive definite, 
cannot be used for proving the $c$-theorem in four dimensions
\cite{cfl}\cite{os}.

\bigskip


\subsection{The three-point function}

The three-point function has the following structure \cite{kaguma}:
\bea
\!\!\! &&\bra T_{\mu_3\nu_3}(-k_1-k_2)\ T_{\mu_1\nu_1}(k_1)\ 
T_{\mu_2\nu_2}(k_2) \ket \nl
\!\!\! && = \sum_{i=1}^{20}\ A_i(q^2,k^2) \ 
{\cal P}^{(i)}_{\mu_3\nu_3,\mu_1\nu_1,\mu_2\nu_2}\ ,\nl
\!\!\! &&\ \  k^2\equiv k_1^2=k_2^2\ , \quad q^\mu\equiv -k_1^\mu-k_2^\mu\ ,
\label{3t4}\eea
where the tensors ${\cal P}^{(i)}$ have dimensions greater or equal to four.
This results has been obtained by solving
the Ward identities for Diffeomorphism invariance starting
from a general expansion involving $137$ basic polynomes;
the involved tensor algebra can be overcomed 
by using algebraic programs.

In Equation (\ref{3t4}), $16$ amplitudes are proper of the three-point
function,  while $4$ are linked to the two-point function.
Two among the $16$ amplitudes match the Euler and Weyl
anomalies at criticality:
\bea
A_E \left(q^2,k^2 \right)\ \ \ \ \  &\to &\ \ \l\ \frac{a}{q^2}\ ,
\qquad\quad   {\rm fixed\ points},\nl
A_W \left(q^2,k^2=0 \right) &\to & - \l\ \frac{3c}{q^2}\ ,
\label{a3-crit}\eea
These limits are obtained by solving the Ward identity
for the Weyl symmetry of the critical theory,
which is anomalous according to (\ref{trace}).
The amplitudes in (\ref{a3-crit}) have dimension $(-2)$ 
because the corresponding tensors are six-dimen\-sional,
and thus are scheme independent.
Equations (\ref{a3-crit}) give the expected relation
between the anomaly coefficients and the scheme-inde\-pendent correlations
in four dimensions; the corresponding two-dimensional
relation is given by Eq.(\ref{2d-c}).

Other scheme-independent amplitudes are non-vanishing at criticality 
(see also the analysis of Ref.\cite{ol});
two further amplitudes of zero dimension account for
the scheme dependence of the three-point function, including
that pertaining to $a'$ \cite{kaguma}.

Each amplitude in the expansion (\ref{3t4})
can be singled out by projecting the three-point function
with the help of the dual tensor basis defined by:
\beq
\left(\ {\cal P}^{*(i)}\ \Big\vert\ {\cal P}^{(j)}\ \right) =
\d^{ij}\ ,
\label{dual}\eeq
where the non-degenerate scalar product is obtained by
contracting the six indices.


\subsection{Results and Conclusions}

Let us now discuss some consequences 
of the general expression (\ref{3t4}) of the three-point function: 

\noindent{\it i)}
It disentangles the kinematic properties of field theory, 
such as Poincar\'e, Weyl and Bose symmetry, from the dynamics 
encoded in the scalar amplitudes.

\noindent{\it ii)}
The imaginary part of any scheme-independent amplitude describes a physical 
quantity such as a scattering or a decay process.

\noindent{\it iii)}
The results (\ref{3t4}, \ref{a3-crit}) amount to a re-derivation of the trace
anomaly within the dispersive renormalization, in close analogy to
the well-know analysis of the chiral triangle $\bra AVV\ket$ 
of Ref.\cite{fsby}; incidentally, the relations (\ref{3t4}, \ref{a3-crit})
can be practically useful for deriving the trace anomaly by Feynman diagram
calculations.

\noindent{\it iv)}
We can write sum rules for the RG flows of
the $a$ and $c$ coefficients in close analogy with the two-dimensional 
case described by Eqs.(\ref{disp}-\ref{s-rule}).
For the $A_E$ amplitude, we write (similar expressions can be written for
$A_W$ at $k^2=0$):
\beq
A_E \left( q^2,k^2 \right) = \int ds\ \frac{\r_E(s,k^2)}{s+q^2}
\ ,
\label{disp-e}\eeq
where the measure $\r_E(s,k^2) ds$ reduces at
criticality to (cf. (\ref{a3-crit})):
\beq
\r_E \left(s, k^2 \right)\ ds \to \l\ a\ \d(s)\ ds\ ,
\qquad {\rm fixed\ points}.
\label{r-crit-e}\eeq

The properties of this measure are very similar to that of its two-dimensional 
counterpart (\ref{disp}): it is a finite dimensionless function of the
renormalized coupling constants (i.e. of the mass scale off-criticality $\ m$),
which satisfies an homogeneous RG equation; note, however,
the dependence on two variables rather then one.

Following the same steps as in Section 3.1, we arrive to
the sum rule:
\bea
a_{UV}-a_{IR} &=& \frac{1}{\l}
\int_\eps^\infty\ ds\ \r_E\left(s, k^2=0 \right) \nl
&=& \frac{1}{\l}\int_\eps^\infty ds\ \Im\TTT\vert_{{\cal P}_E}\ \ 
\label{a-rule}\eea
In this Equation, we have set the second momentum $k^2$ to zero,
in order to let the measure to depend on the ratio $\ s/m^2\ $ only. 

Note that the sum rule (\ref{a-rule}) is not enough to prove
$a_{UV} > a_{IR}$, 
because the $\ \r_E\ $ measure is not manifestly positive definite.
A positivity condition for the three-point function
has been proposed in Ref.\cite{ol}, following from 
the (quantum) weak-energy condition, but its consequences on $\r_E$
and $\r_W$ remain to be explored.
Manifestly positive amplitudes occur in the four-point function,
which could also be analysed using the same tools.

In conclusion, we hope that the general expansion (\ref{3t4})
and its dispersive analysis will be useful for further investigations.

\acknowledgments
The authors thank the organizers of this Conference for
the opportunity to present this work; we also thank all our
partners of this European network for the many occasions of
informal discussion and collaboration.
This work is supported in part by the European Community  
Network grant FMRX-CT96-0012.


\def\NPB#1#2#3{{\it Nucl.~Phys.} {\bf{B#1}} (#2) #3} 
\def\CMP#1#2#3{{\it Commun.~Math.~Phys.} {\bf{#1}} (#2) #3} 
\def\CQG#1#2#3{{\it Class.~Quantum~Grav.} {\bf{#1}} (#2) #3} 
\def\PLB#1#2#3{{\it Phys.~Lett.} {\bf{B#1}} (#2) #3} 
\def\PRD#1#2#3{{\it Phys.~Rev.} {\bf{D#1}} (#2) #3} 
\def\PRL#1#2#3{{\it Phys.~Rev.~Lett.} {\bf{#1}} (#2) #3} 
\def\ZPC#1#2#3{{\it Z.~Phys.} {\bf C#1} (#2) #3} 
\def\PTP#1#2#3{{\it Prog.~Theor.~Phys.} {\bf#1}  (#2) #3} 
\def\MPLA#1#2#3{{\it Mod.~Phys.~Lett.} {\bf#1} (#2) #3} 
\def\PR#1#2#3{{\it Phys.~Rep.} {\bf#1} (#2) #3} 
\def\AP#1#2#3{{\it Ann.~Phys.} {\bf#1} (#2) #3} 
\def\RMP#1#2#3{{\it Rev.~Mod.~Phys.} {\bf#1} (#2) #3} 
\def\HPA#1#2#3{{\it Helv.~Phys.~Acta} {\bf#1} (#2) #3} 
\def\JETPL#1#2#3{{\it JETP~Lett.} {\bf#1} (#2) #3} 
\def\JHEP#1#2#3{{\it JHEP} {\bf#1} (#2) #3} 
\def\TH#1{{\tt hep-th/#1}}

\end{document}